\documentclass[twocolumn,american,nofootinbib,floatfix]{revtex4-1}
\usepackage[T1]{fontenc}
\usepackage[utf8]{inputenc}
\usepackage{amsmath}
\usepackage{graphicx}

\makeatletter

\providecommand{\tabularnewline}{\\}

\usepackage{color}

\usepackage[english]{babel}

\providecommand{\tabularnewline}{\\}

\usepackage[caption=false]{subfig}

\usepackage{babel}

\@ifundefined{showcaptionsetup}{}{%
 \PassOptionsToPackage{caption=false}{subfig}}
\usepackage{subfig}
\makeatother

\usepackage{babel}
\begin{document}
\title{Dynamical manifestation of Gibbs paradox after a quantum quench}
\author{M. Collura$^{1}$, M. Kormos$^{2}$ and G. Takács$^{2,3}$}
\thanks{Corresponding author (email: takacsg@eik.bme.hu)}
\affiliation{$^{1}$The Rudolf Peierls Centre for Theoretical Physics, ~~\\
 Oxford University, Oxford, OX1 3NP, United Kingdom~~\\
 $^{2}$BME ``Momentum'' Statistical Field Theory Research Group,
H-1117 Budapest, Budafoki út 8.~~\\
 $^{3}$BME Department of Theoretical Physics, H-1117 Budapest, Budafoki
út 8.}
\date{29th September 2018}
\begin{abstract}
We study the propagation of entanglement after quantum quenches in
the non-integrable paramagnetic quantum Ising spin chain. Tuning the
parameters of the system, we observe a sudden increase in the entanglement
production rate, which we show to be related to the appearance of
new quasi-particle excitations in the post-quench spectrum. We argue
that the phenomenon is the non-equilibrium version of the well-known
Gibbs paradox related to mixing entropy and demonstrate that its characteristics
fit the expectations derived from the quantum resolution of the paradox
in systems with a non-trivial quasi-particle spectrum. 
\end{abstract}
\maketitle

\section{Introduction}

A quantum quench is a protocol routinely engineered in cold-atom experiments
\cite{exp1,exp2,exp3,exp3a,exp4,exp5,exp6,exp7,exp8}: a sudden change
of the Hamiltonian of an isolated quantum system followed by a non-equilibrium
time evolution. The initial state corresponds to a highly excited
configuration of the post-quench Hamiltonian, acting as a source of
quasi-particle excitations \cite{calabrese-cardy}. In a large class
of systems, there is a maximum speed for these excitations called
the Lieb-Robinson bound \cite{lieb-robinson} which results in a linear
growth of entanglement entropy $S(t)\sim t$ of a subsystem of length
$\ell$ for times $t<\ell/2v_{\text{max}}$, after which it becomes
saturated \cite{calcard_entropy}. The mean entropy production rate
$\overline{\partial_{t}S}$ characterizing the linear growth naturally
depends on the post-quench spectrum and reflects its quasi-particle
content.

Entanglement entropy contains a wealth of information regarding the
non-equilibrium evolution and the stationary state resulting after
a quench, and therefore has been studied extensively in recent years
\cite{ent1,ent2,ent3,ent4,ent6,ent7,ent8,ent9,ent10,ent11,ent12,ent13,ent14,ent15}.
The growth of entanglement also has important implications for the
efficiency of computer simulations of the time evolution \cite{sim1,sim3,sim4,sim5}.
Recently it has become possible to measure entanglement entropy and
its temporal evolution in condensed matter systems \cite{exp2,Smeas1,Smeas2}.
For integrable systems, an analytic approach of entanglement entropy
production has been developed recently in \cite{integrable_entanglement_growth,integrable_renyi,multi-particle_case}. 

In this paper we consider quenches in the quantum Ising chain by switching
on an integrability breaking longitudinal magnetic field $h_{x}$
in the paramagnetic phase. In similar quenches in the ferromagnetic
regime, it was recently found that confinement suppresses the usual
linear growth of entanglement entropy and the corresponding light-cone-like
spreading of correlations after the quantum quench \cite{Ising_confinement}.
However, in the paramagnetic regime considered here confinement is
absent and thus entanglement entropy grows linearly in time. Nevertheless
the dependence of the entropy production rate on the quench parameter
$h_{x}$ shows another kind of anomalous behavior: a sudden increase
setting in at the threshold value of $h_{x}$ where a new quasi-particle
excitation appears in the spectrum.

Using the physical interpretation of the asymptotic entanglement of
a large subsystem as the thermodynamic entropy of the stationary (equilibrium)
state \cite{exp2,calcard_entropy,integrable_entanglement_growth,deutsch,beugeling},
this can be recognized as arising from the contribution of mixing
entropy between the particle species, and therefore constitutes a
non-equilibrium manifestation of the Gibbs paradox.

\section{Entropy production rate as a function of the longitudinal field}

\begin{figure*}[t]
\begin{centering}
\includegraphics[width=0.8\paperwidth]{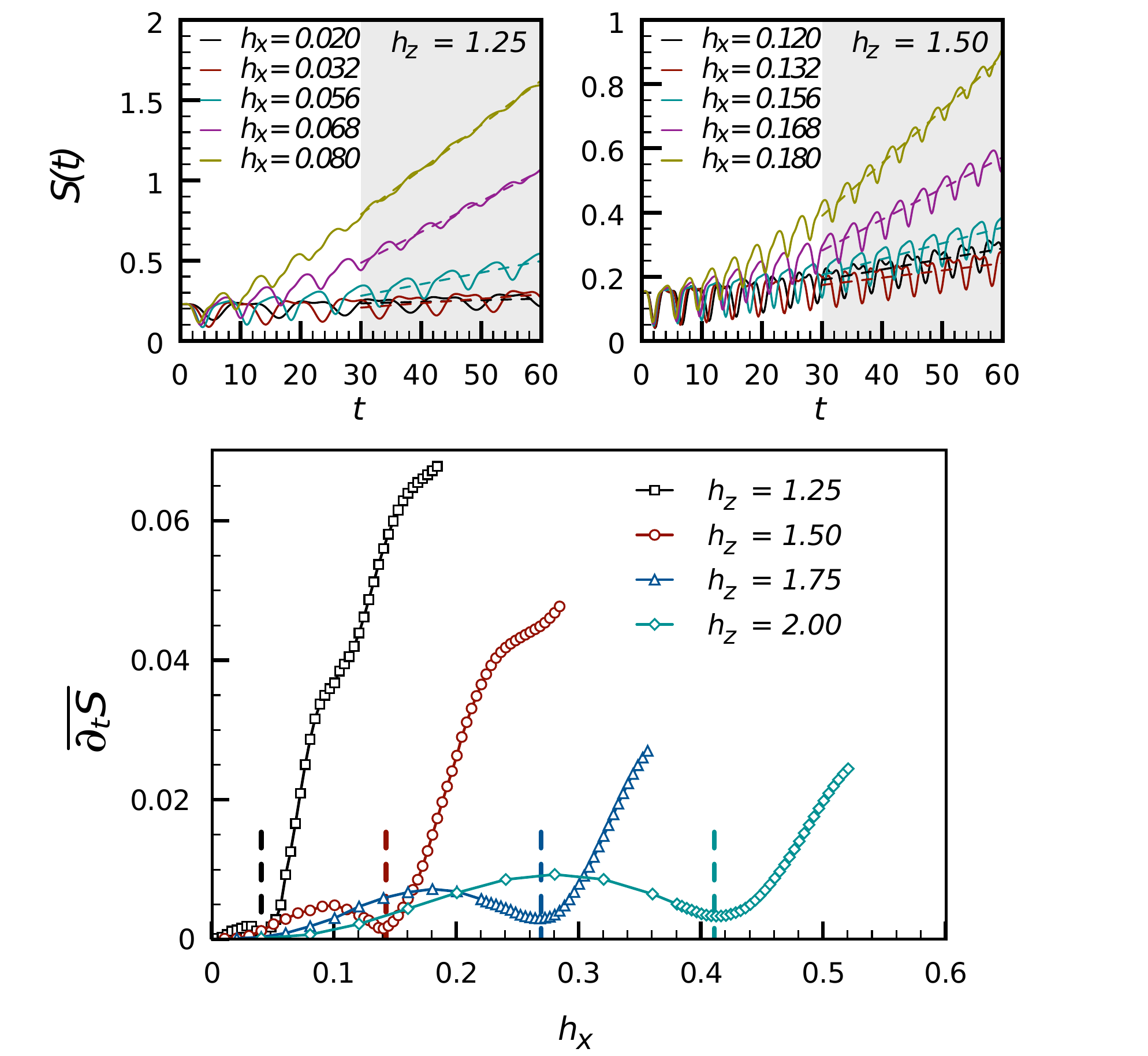} 
\par\end{centering}
\caption{\label{fig:Top-diagrams:-time} Top panel: time dependence of half-system
entanglement entropy as the post-quench longitudinal field $h_{x}$
is changed. The shaded regions show the time interval used to fit
the linear time dependence. Bottom panel: the mean entropy production
rate $\overline{\partial_{t}S}$, defined as the slope of the linear
part of the entanglement entropy $S(t)$, is shown as a function of
$h_{x}$, with the dashed vertical lines showing the position of the
minimum $h_{x}^{\text{min}}$.}
\end{figure*}
The Ising quantum spin chain is defined by the Hamiltonian 
\begin{equation}
H=J\sum_{i=0}^{L-1}\left(-\sigma_{i}^{x}\sigma_{i+1}^{x}+h_{z}\sigma_{i}^{z}+h_{x}\sigma_{i}^{x}\right)\;,
\end{equation}
where $\sigma_{i}^{x,z}$ denote the standard Pauli matrices acting
at site $i$, and we assume periodic boundary conditions $\sigma_{L}^{x,z}\equiv\sigma_{0}^{x,z}$. 

It is exactly solvable for $h_{x}=0$ with a quantum critical point
at $h_{z}=1$. For $h_{z}<1$, the system shows ferromagnetic ordering
with order parameter $\sigma_{i}^{x}$ . The paramagnetic phase corresponds
to transverse magnetic field $h_{z}>1$, where the spectrum consists
of free fermionic excitations over a unique ground state with the
dispersion relation 
\begin{align}
\epsilon(k_{n}) & =2J\sqrt{1+h_{z}^{2}-2h_{z}\cos k_{n}}\;,\label{eq:integrable_disprel}\\
 & k_{n}=n\frac{2\pi}{L}\ ,\ n=-\frac{L}{2}+1,-\frac{L}{2}+2,\dots,\frac{L}{2}\,,\nonumber 
\end{align}
where we assumed that the chain length $L$ is even. The fermionic
quasi-particles correspond to spin waves with the maximum propagation
velocity of $(d\epsilon/dk)_{\text{max}}=2J$.

We consider quantum quenches in the thermodynamic limit $L\rightarrow\infty$.
We prepare the system in the ground state $|\Psi_{0}\rangle$ at some
$h_{z}>1$ and $h_{x}=0$, and study the time evolution of the half-system
entanglement after suddenly switching on a non-zero $h_{x}$ leading
to the time-dependent state $|\Psi(t)\rangle=\exp(-iHt)|\Psi_{0}\rangle$
using the infinite size time evolving block decimation (iTEBD) algorithm
\cite{vidal1}.

The standard measure of the half-system entanglement is \cite{entS1,entS2,entS3}
\begin{equation}
S(t)=-\text{Tr}_{R}\rho_{R}(t)\log\rho_{R}(t)\;,
\end{equation}
which is just the von Neumann entropy of the reduced density matrix
$\rho_{R}(t)=\text{Tr}_{L}|\Psi(t)\rangle\langle\Psi(t)|$ of one
half of the system obtained by tracing out the other half. The entropy
production rate obtained from iTEBD simulations is shown in Fig. \ref{fig:Top-diagrams:-time}.

The top panel demonstrates that the average late time behavior of
the entanglement entropy $S(t)$ can be fit with a linear behavior
apart from slowly decaying periodic fluctuations, as expected after
a global quantum quench \cite{calcard_entropy}. The mean entanglement
production rate $\overline{\partial_{t}S}$ is obtained from the slope
of the linear part and can be interpreted as the production rate of
the thermodynamic entropy. In the bottom panel the dependence of $\overline{\partial_{t}S}$
on $h_{x}$ is shown. After some initial increase the entropy production
rate starts decreasing, but at some value of the longitudinal field
$h_{x}$ this trend gets reversed in a dramatic fashion and turns
into a rapid increase. This surprising trend change happens at the
value $h_{x}^{\mathrm{min}}$ where $\overline{\partial_{t}S}$ has
a local minimum; the measured positions of these minima are listed
in Table \ref{tab:Position-of-the}. As we demonstrate below, the
explanation of this curious behavior lies in the quasiparticle content
of the model.
\begin{center}
\begin{table}
\begin{centering}
\begin{tabular}{|c|c|c|c|c|}
\hline 
$h_{z}$  & $1.25$  & $1.5$  & $1.75$  & $2$\tabularnewline
\hline 
\hline 
$h_{x}^{\text{min}}$  & $0.040$  & $0.140$  & $0.268$  & $0.412$\tabularnewline
\hline 
\end{tabular}
\par\end{centering}
\caption{\label{tab:Position-of-the}Position of the local minimum of $\overline{\partial_{t}S}$
as a function of $h_{x}$ for different values of $h_{z}$.}
\end{table}
\par\end{center}

\section{Quasi-particle spectrum of the paramagnetic Ising chain\label{sec:Quasi-particle-spectrum-of}}

Switching on the longitudinal field $h_{x}\neq0$ breaks integrability,
but the spectrum can be determined by numerical methods by exact diagonalization
which we applied to chains of length $L=16,\,18,\,20$ and $22$,
using units $J=1$. The energy eigenstate basis can be chosen to be
a simultaneous eigenstate basis of the position shift operator $\mathcal{S}$
defined by
\[
\mathcal{S}\sigma_{i}^{a}\mathcal{S}^{-1}=\sigma_{i+1}^{a}
\]
The eigenvalue of $\mathcal{S}$ is a complex phase $e^{ik}$ where
$k$ is the momentum of the state, defined modulo $2\pi$.

\subsection{The first quasi-particle excitation \label{subsec:The-first-quasi-particle}}

\begin{figure}
\begin{centering}
\includegraphics[width=0.8\columnwidth]{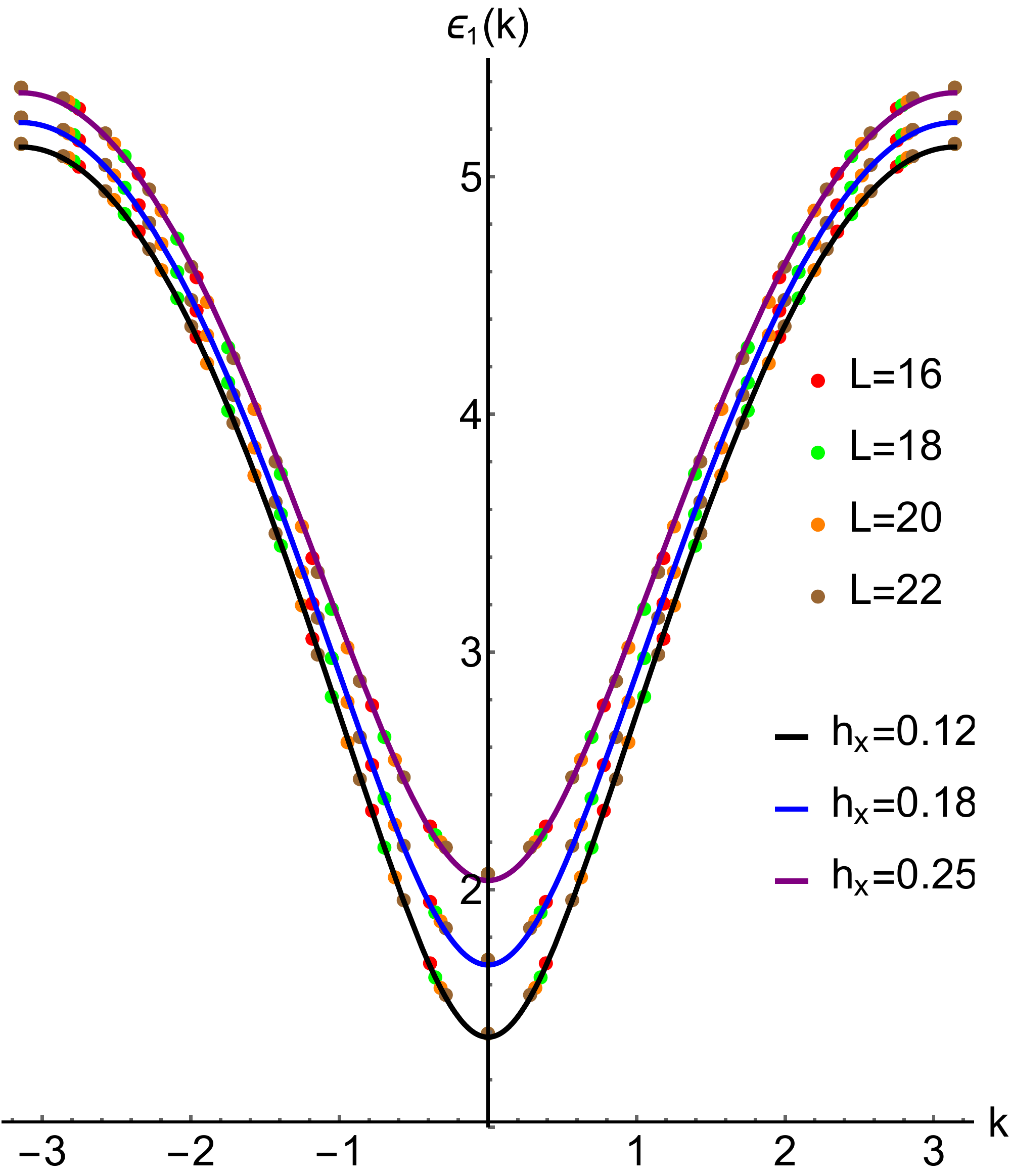} 
\par\end{centering}
\caption{\label{fig:The-first-quasi-particle}The first quasi-particle dispersion
relation for $h_{z}=1.5$ and $h_{x}=0.12$, $0.18$ and $0.25$.
The differently colored dots are energy levels computed for systems
sizes of $L=16$, $18$, $20$ and $22$ spins, illustrating that
finite size dependence is already negligible. The continuous lines
are fits of a function $\epsilon_{1}(k)=\sqrt{A+B\cos k}$.}
\end{figure}

From the numerically computed spectrum, the lowest-lying one-particle
states can be selected as the lowest energy states among those with
a fixed momentum $k\neq0$, while at $k=0$ the relevant state is
the first excited above the ground state; this gives the first quasi-particle
branch. The dispersion relation $\epsilon_{1}(k)$ of the first quasi-particle
can be obtained by subtracting the ground state value, with the result
shown in Fig. \ref{fig:The-first-quasi-particle} for the case $h_{z}=1.5$
and a few values of $h_{x}$. The data can be fitted to a very good
precision with a curve of the form 
\begin{equation}
\epsilon_{1}(k)=\sqrt{A+B\cos k}\;,
\end{equation}
inspired by the exact dispersion relation of the $h_{x}=0$ chain.
It is already apparent from the graph that the quasi-particle gap
(mass) increases with $h_{x}$, while the Lieb-Robinson velocity 
\begin{equation}
v_{\text{max}}=\max_{k}\frac{d\epsilon_{1}}{dk}
\end{equation}
decreases, which can also be shown by computing $v_{\text{max}}$
numerically from the fit with the results shown in Table \ref{tab:Values-of-the}.
\begin{center}
\begin{table}
\begin{centering}
\begin{tabular}{|c|c|c|c|}
\hline 
$h_{x}$  & $0.12$  & $0.18$  & $0.25$\tabularnewline
\hline 
\hline 
$v_{\text{max}}$  & $1.873$  & $1.772$  & $1.657$\tabularnewline
\hline 
\end{tabular}
\par\end{centering}
\caption{\label{tab:Values-of-the}Values of the Lieb-Robinson velocity determined
from the data shown in Fig. \ref{fig:The-first-quasi-particle}.}
\end{table}
\par\end{center}

\begin{figure}
\subfloat[First quasi-particle gap]{\begin{centering}
\includegraphics[width=0.7\columnwidth]{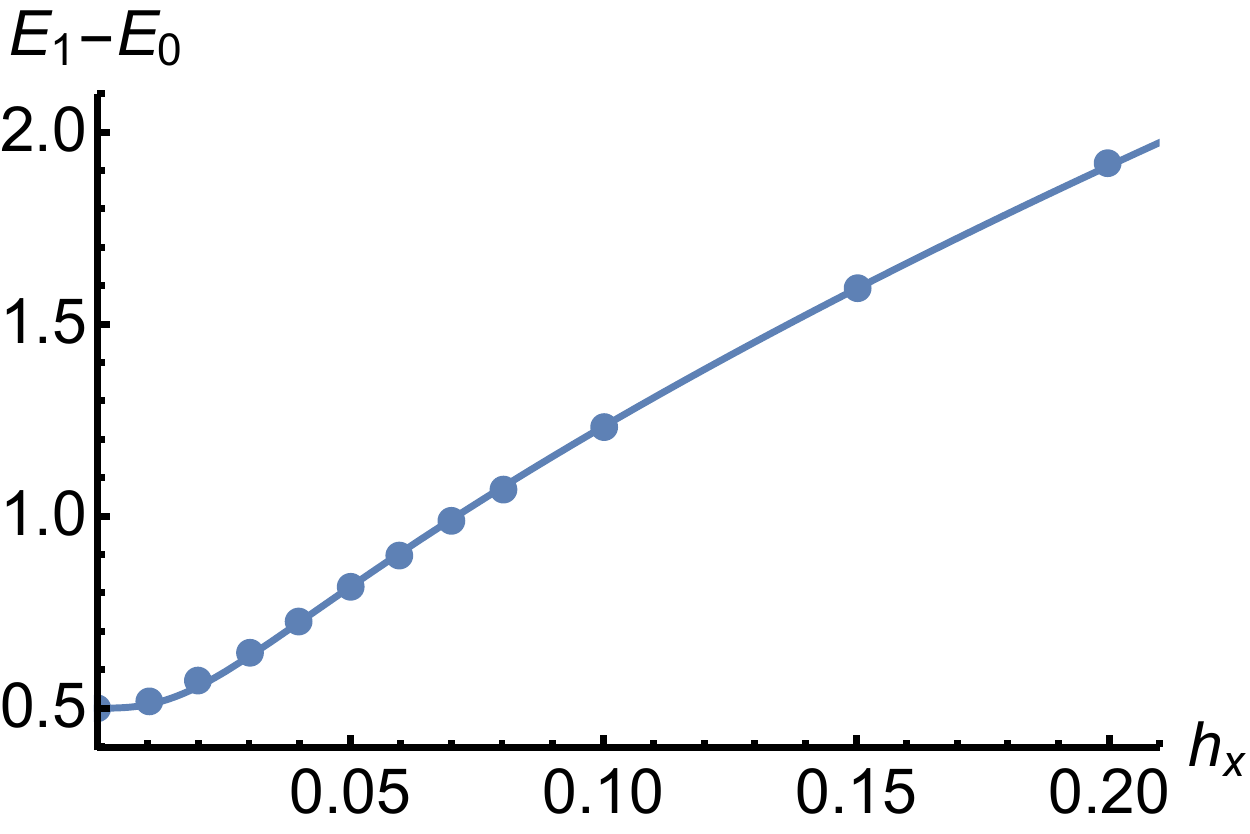} 
\par\end{centering}
}

\subfloat[Lieb-Robinson velocity]{\begin{centering}
\includegraphics[width=0.7\columnwidth]{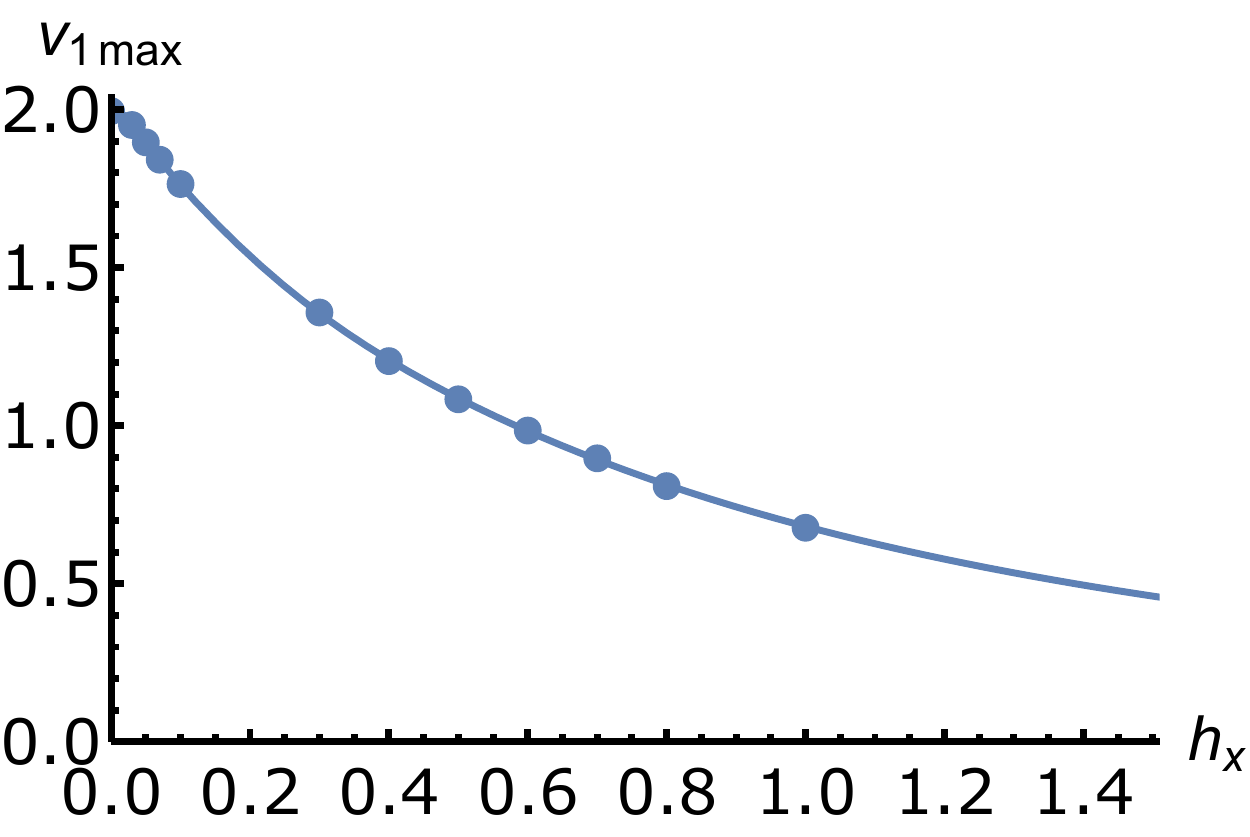} 
\par\end{centering}
}

\caption{\label{fig:The-gap-and}The gap and maximum velocity of the first
quasi-particle as a function of $h_{x}$ for $h_{z}=1.25$.}
\end{figure}

A more complete picture of the properties of the first quasi-particle
is shown in Fig. \ref{fig:The-gap-and} for $h_{z}=1.25$ . This value
was simply chosen for illustration; the qualitative picture does not
change for other values for $h_{z}$. Note that the quasi-particle
mass gets corrections of order $h_{x}^{2}$ for small $h_{x}$ and
becomes linear for large $h_{x}$. The first one can easily be confirmed
by perturbation theory, while the second is a simple consequence of
the form of the Hamiltonian. Also note that the Lieb-Robinson velocity
decreases with increasing $h_{x}$ and eventually goes to zero for
very large $h_{x}$; this is easy to understand since for very large
$h_{x}$ the dynamics of the spins essentially becomes frozen.

\subsection{Bound states in the continuum limit \label{subsec:Bound-states-in}}

In the vicinity of the quantum critical point $h_{z}\sim1$ and $h_{x}\sim0$
it is possible to take a continuum limit to the scaling Ising field
theory. For vanishing $h_{x}$ it describes a massive free Majorana
fermion with mass $M=2J|1-h_{z}|$. For non-zero $h_{x}$, the coupling
corresponding to $h_{x}$ in the continuum limit scales as $h\propto h_{x}J^{15/8}$
\cite{RMCKT}. The scaling limit is obtained by taking $J\rightarrow\infty$
while $h_{z}\rightarrow1$ and $h_{x}\rightarrow0$ such that 
\begin{align}
M & =2J\left|1-h_{z}\right|\,,\nonumber \\
h & =\frac{2}{\bar{s}}J^{15/8}\,h_{x}\,,\quad\bar{s}=2^{1/12}e^{-1/8}\mathcal{A}^{3/2}\nonumber \\
 & \mathcal{A}=1.282427129\dots
\end{align}
are kept fixed, the quantum Ising spin chain scales to the Ising field
theory given in terms of a Majorana fermion field $\psi,\bar{\psi}$
\begin{align}
H_{\text{IFT}} & =\int_{-\infty}^{\infty}dx\,\Bigg\{\frac{1}{2\pi}\Bigg[\frac{i}{2}\left(\psi(x)\partial_{x}\psi(x)-\bar{\psi}(x)\partial_{x}\bar{\psi}(x)\right)\nonumber \\
 & \qquad-iM\bar{\psi(}x)\psi(x)\Bigg]+h\sigma(x)\Bigg\}\,,\nonumber \\
 & \left\{ \psi(x,t),\bar{\psi}(y,t)\right\} =2\pi\delta(x-y)\:,
\end{align}
using units in which the lattice spacing is $a=2/J$ and the resulting
speed of light is $c=1.$ The operator $\sigma(x)$ is the continuum
limit of magnetization $\sigma_{i}^{x}$ which is non-local with respect
to the Majorana fermionic field and corresponds to a twist field changing
the boundary condition of the fermion from periodic to anti-periodic
and vice versa.

A detailed numerical study of the field theory limit revealed that
switching on a longitudinal field $h$ leads to the appearance of
a second and a third quasi-particle excitation at some threshold values
$h_{c1}$ and $h_{c2}$ which scale as $M^{15/8}$ \cite{zam_ising_spectr}.
These excitations can be considered bound states of the fundamental
one, and the spectrum only depends on the dimensionless ratio $\chi=M/h^{8/15}$,
with $h=0$ corresponding to $\chi=\infty$.

One can also approach the question of spectrum from the other extremal
point $\chi=0$, that is the case of $M=0$ when one obtains the famous
$E_{8}$ model \cite{e8}. At this point there exist $8$ particles
with masses $m_{i}$ in the continuum limit, the ratios of which are
known exactly, with the first two having the values 
\begin{align}
\Delta_{21} & =\frac{m_{2}}{m_{1}}=2\cos\frac{\pi}{5}=1.618\dots\nonumber \\
\Delta_{31} & =\frac{m_{2}}{m_{1}}=2\cos\frac{\pi}{30}=1.989\dots
\end{align}
As soon as one switches on a mass $M$ which takes the system into
the paramagnetic regime\footnote{In fact, this is a little more complicated as the sign of mass term
is irrelevant in the field theory. In the continuum limit, the distinction
between the two phases is encoded in the Hilbert space, cf. Ref. \cite{RMCKT}.} (corresponding to $h_{z}>1$), all but three of these particles become
unstable \cite{zam_ising_spectr}. Further increasing $M$ (more precisely,
the dimensionless ratio $\chi$) makes the third particle unstable
in short order, with the second particle disappearing for much larger
values of $\chi$ \cite{zam_ising_spectr}. For the limit of infinite
$\chi$ which corresponds to $h=0$ i.e. a free massive Majorana fermion,
only a single particle remains in the spectrum.

\subsection{The bound state quasi-particles on the chain \label{subsec:The-bound-state}}

Turning to the spin chain, now we demonstrate that the quasi-particle
spectrum obtained in the scaling limit persists also for finite lattice
spacing. For a fixed value of $h_{z}$ there exists a threshold value
$h_{x}^{(2)}$ at which a new quasi-particle appears in the spectrum
which can be identified as a bound state of the fundamental quasi-particle,
as in the field theory. For values of $h_{z}$ close enough to the
critical point ($h_{z}=1$) a third quasi-particle can also be found
at sufficiently high $h_{x}$ with a threshold value $h_{x}^{(3)}$;
however, this excitation is always very weakly bound.

The lowest branch of excitations discussed in Subsection \ref{subsec:The-first-quasi-particle}
correspond to the first quasi-particle, and for small enough $h_{x}$
the excitations just above the first quasi-particle branch can be
interpreted as two-particle states. However, for $h_{x}>h_{x}^{(2)}$
the gap to the second branch drops below twice the value of the first
quasi-particle gap, which signals the appearance of stable bound states
forming a second quasi-particle branch. For even higher values $h_{x}>h_{x}^{(3)}$
another branch drops below twice the first gap, signaling the presence
of the third quasi-particle excitation in the spectrum.

To find the bound state thresholds $h_{x}^{(a)}$ ($a=2,3$) above
which the new quasi-particles appear, we took the first four zero-momentum
eigenvalues at chain length $L$ ordered as $E_{0}(L)<E_{1}(L)<E_{2}(L)<E_{3}(L)$,
and computed the gap ratios 
\begin{equation}
\Delta_{21}(L)=\frac{E_{2}(L)-E_{0}(L)}{E_{1}(L)-E_{0}(L)}\qquad\Delta_{31}(L)=\frac{E_{3}(L)-E_{0}(L)}{E_{1}(L)-E_{0}(L)}\;,\label{eq:finvol_gap_ratio}
\end{equation}
which were then extrapolated in $L$ using 
\begin{equation}
\Delta_{a1}(L)=\Delta_{a1}+\gamma_{a1}e^{-\mu_{a1}L}\quad a=2,3\,.\label{eq:extrapol_gap_ratio}
\end{equation}
The condition for the existence of the bound states $a=2,3$ is that
their decay is kinematically forbidden, i.e. $\Delta_{a1}<2$, since
the model is non-integrable and there are no conserved charges to
prevent their decay.

The exponential volume dependence is expected to be valid when the
bound state exists \cite{luscher_onept}, so the extrapolation was
performed in the regime when $h_{x}$ approaches the critical value
$h_{x}^{(a)}$ from above. The exponent $\mu_{a1}$ is related to
the spatial extension of the bound state wave function, while $\gamma_{a1}$
is the interaction strength between the constituents which is negative
as long as the bound state exists, i.e. above the critical value corresponding
to an attractive interaction. Below the critical value, the energy
level corresponds to a two-particle scattering state which is expected
to have power-like leading finite size corrections \cite{luscher_2particle}.
Despite this, the numerical fit with the exponential dependence works
quite well close to the threshold value $h_{x}^{(a)}$ and confirms
the change of the sign in $\gamma_{a1}$ which corresponds to the
interaction becoming repulsive.

\begin{figure*}
\subfloat[$h_{z}$=1.25]{\begin{centering}
\includegraphics[width=0.4\textwidth]{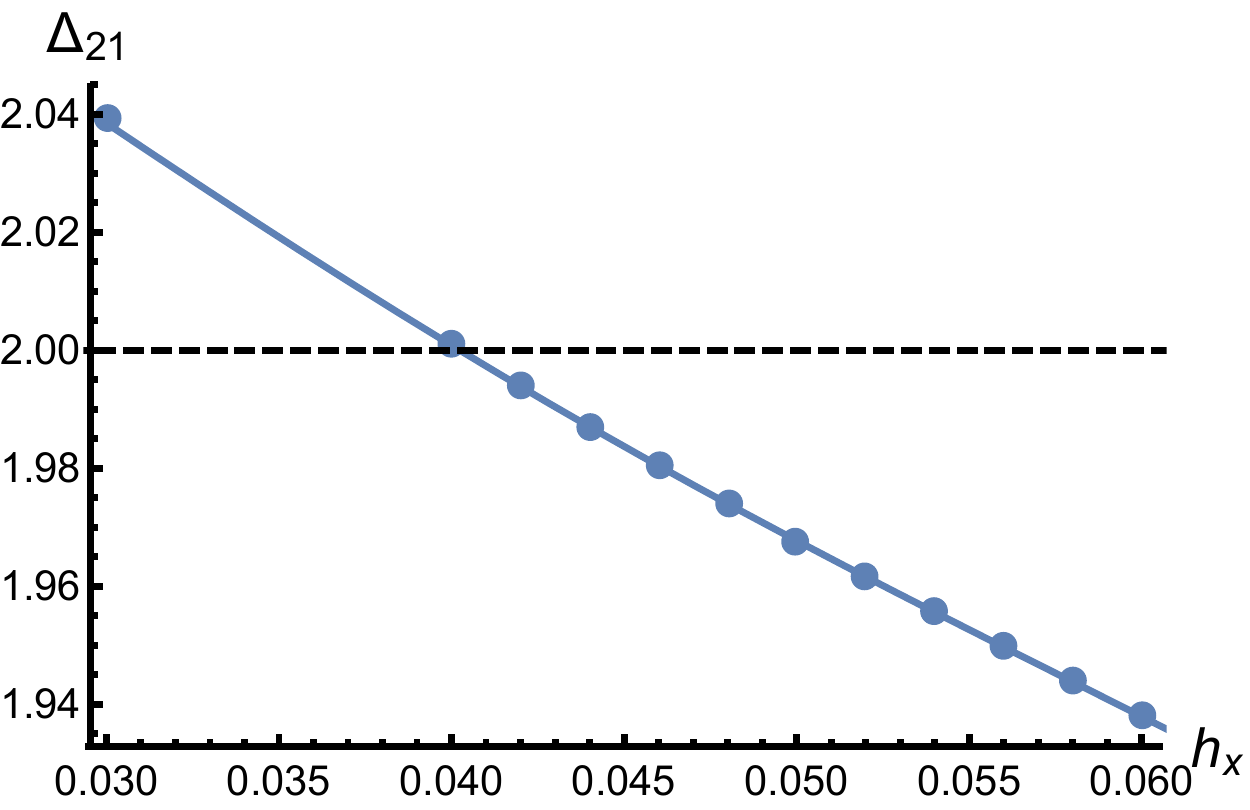} 
\par\end{centering}
}\subfloat[$h_{z}$=1.5]{\begin{centering}
\includegraphics[width=0.4\textwidth]{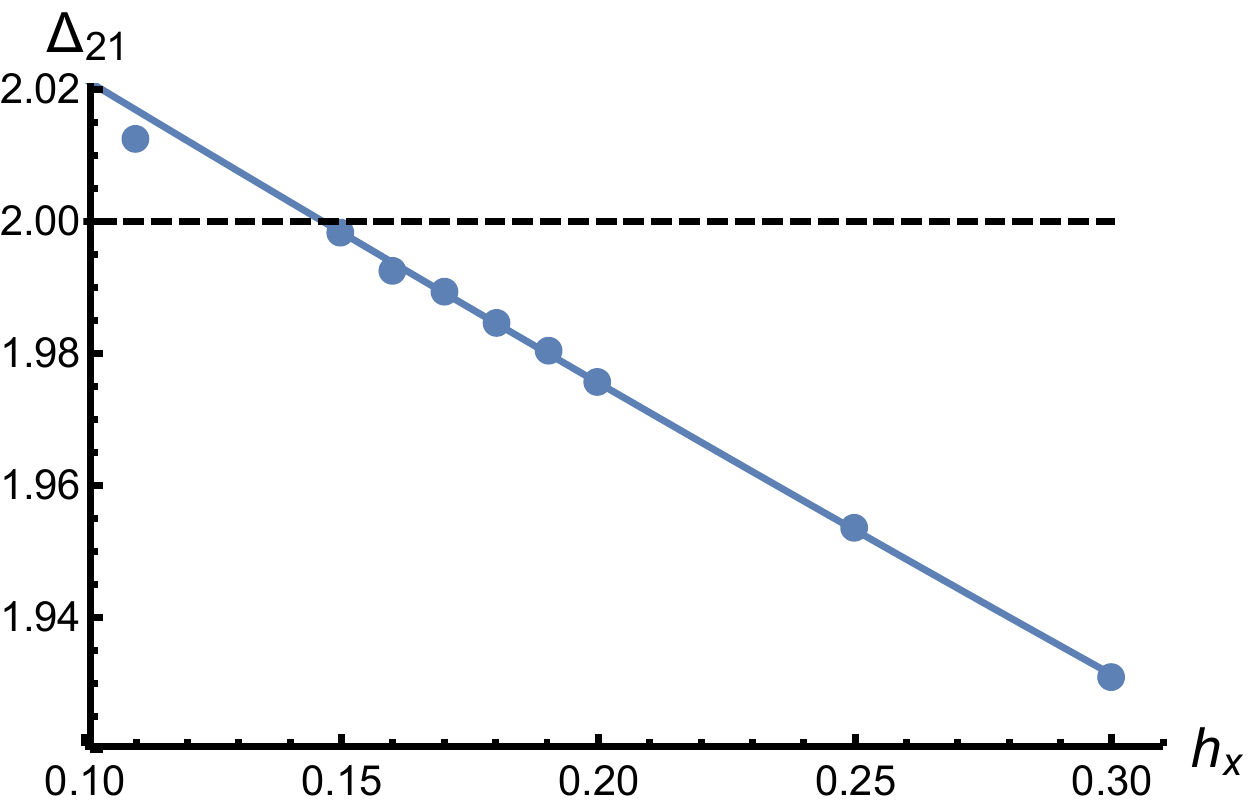} 
\par\end{centering}
}

\subfloat[$h_{z}$=1.75]{\begin{centering}
\includegraphics[width=0.4\textwidth]{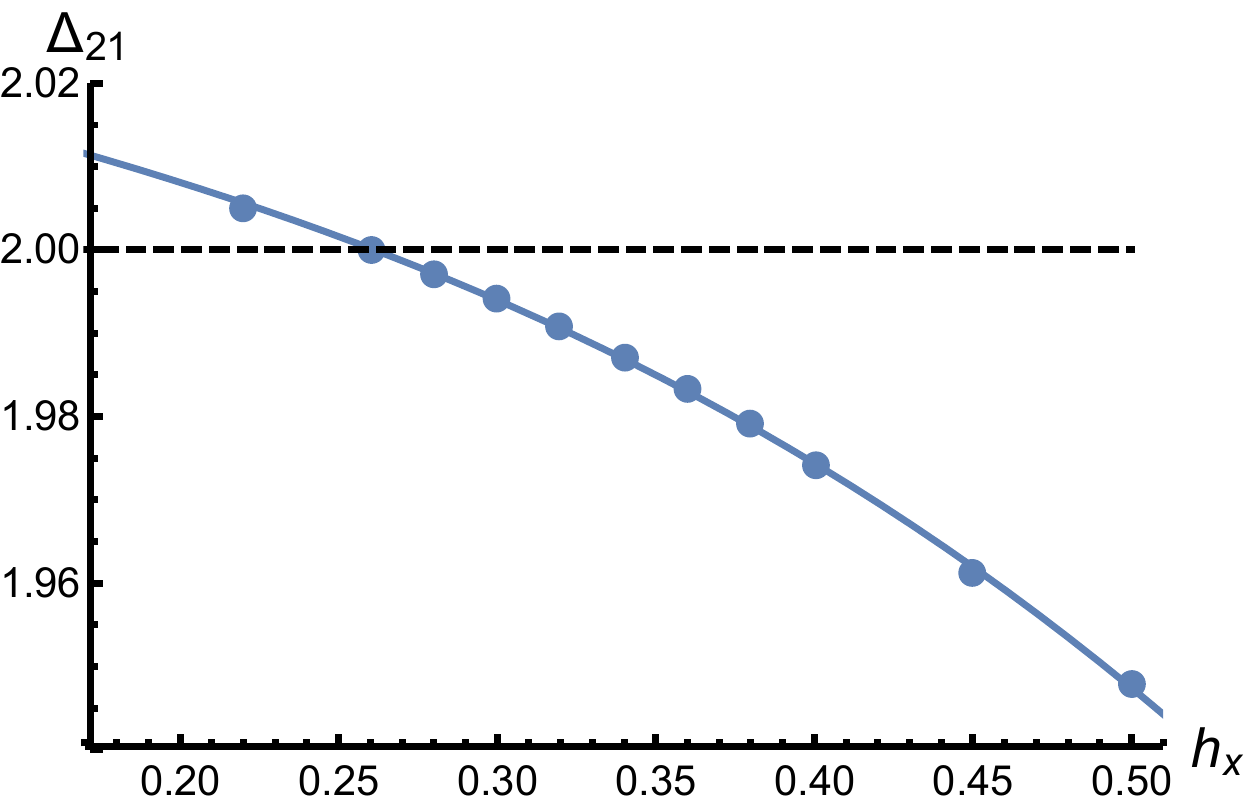} 
\par\end{centering}
}\subfloat[$h_{z}$=2]{\begin{centering}
\includegraphics[width=0.4\textwidth]{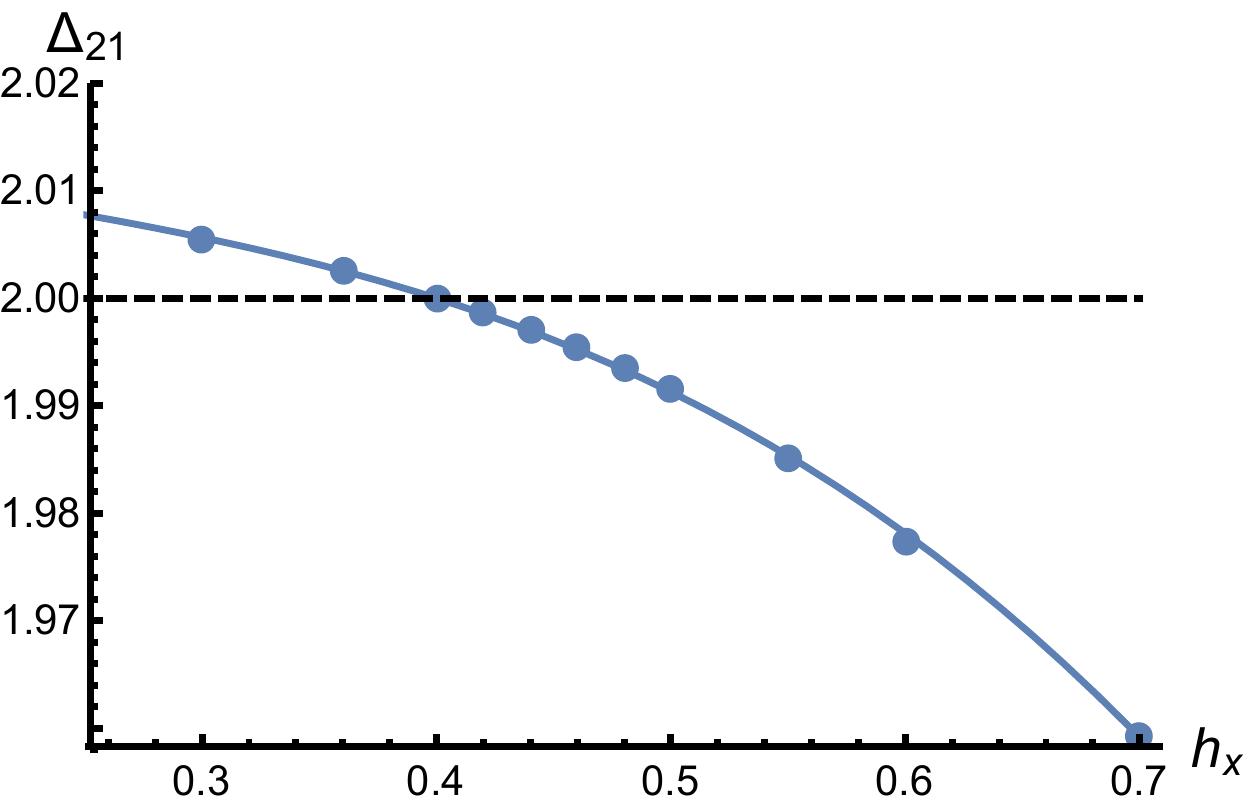} 
\par\end{centering}
}

\caption{\label{fig:Delta12-hx}Gap ratio $\Delta_{21}$ defined in \eqref{eq:extrapol_gap_ratio}
as a function of $h_{x}$.}
\end{figure*}

For four different values of $h_{z}=1.25,\,1.5,\,1.75$ and $2$,
the critical values $h_{x}^{(a)}$ where a given bound state appears
were found numerically from the condition $\Delta_{a1}=2$ as illustrated
in Fig. \ref{fig:Delta12-hx}. The critical values determined numerically
are given in Table \ref{tab:Critical-values-of}.
\begin{center}
\begin{table}
\begin{centering}
\begin{tabular}{|c|c|c|c|c|}
\hline 
$h_{z}$  & $1.25$  & $1.5$  & $1.75$  & $2$\tabularnewline
\hline 
\hline 
$h_{x}^{(2)}$  & $0.040$  & $0.146$  & $0.261$  & $0.400$\tabularnewline
\hline 
\end{tabular}
\par\end{centering}
\caption{\label{tab:Critical-values-of}Critical values of $h_{x}$ corresponding
to the bound state threshold at some values of $h_{z}$.}
\end{table}
\par\end{center}

\begin{center}
\begin{figure*}
\begin{centering}
\subfloat[$h_{z}=1.75$]{\begin{centering}
\includegraphics[width=0.4\textwidth]{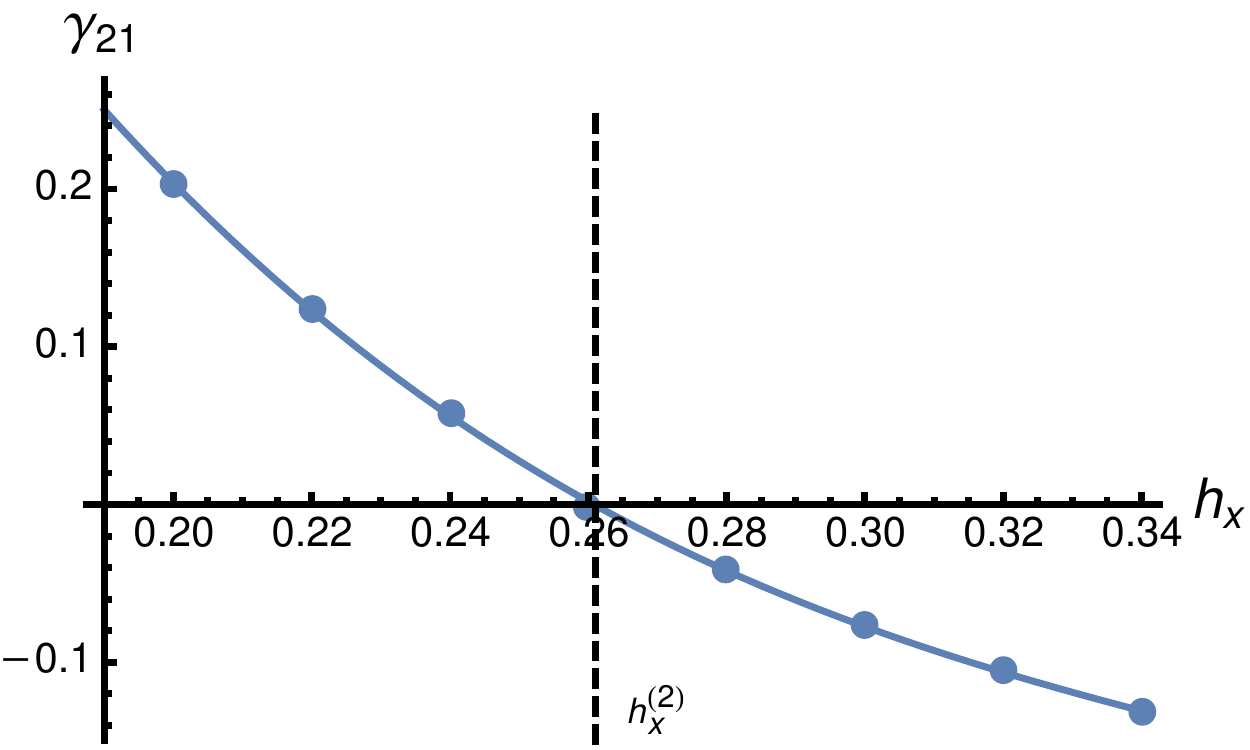} 
\par\end{centering}
}\subfloat[$h_{z}=2$]{\begin{centering}
\includegraphics[width=0.4\textwidth]{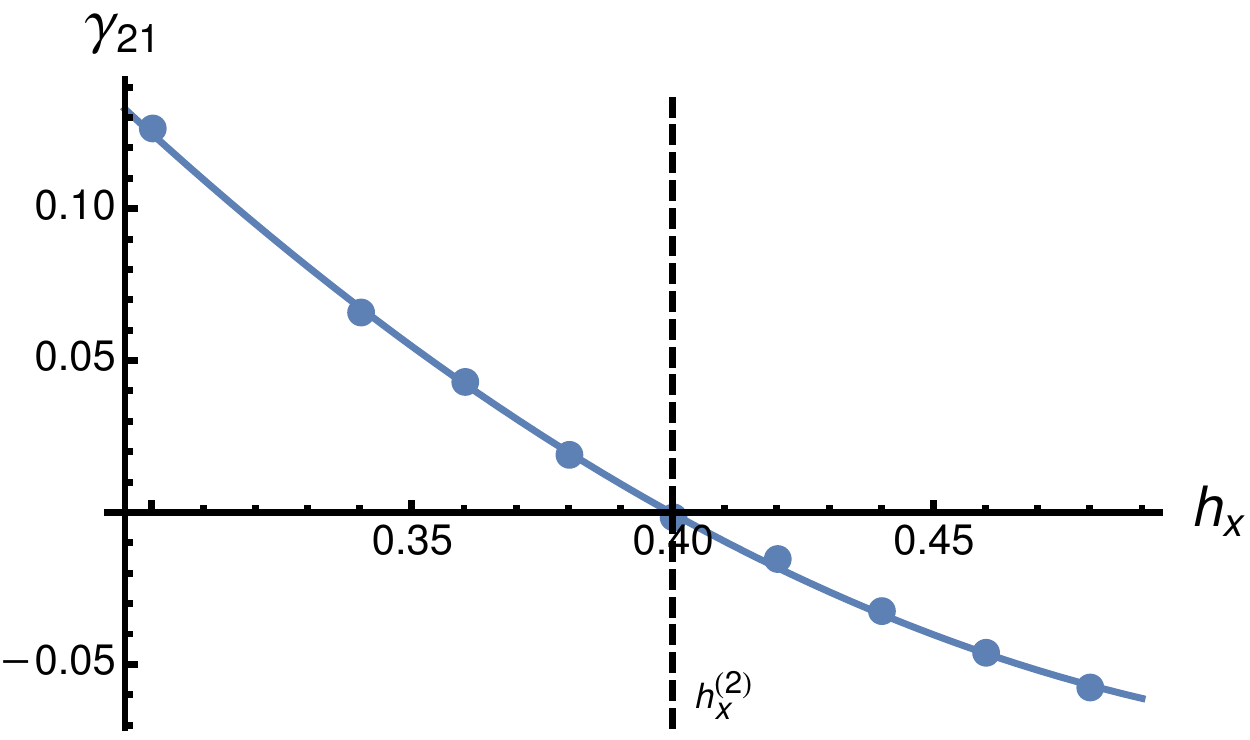} 
\par\end{centering}
}
\par\end{centering}
\caption{\label{fig:Dependence-of-gamma21-on-hx}Interaction strength $\gamma_{21}$
defined in \eqref{eq:extrapol_gap_ratio} as a function of $h_{x}$. }
\end{figure*}
\par\end{center}

In Fig. \ref{fig:Dependence-of-gamma21-on-hx} it is demonstrated
using the examples of $h_{z}=1.75$ and $2$ that $\gamma_{21}$ indeed
changes sign at $h_{x}^{(2)}$ extracted from the above threshold
criterion $\Delta_{21}=2$.

\begin{figure}
\begin{centering}
\includegraphics[width=0.8\columnwidth]{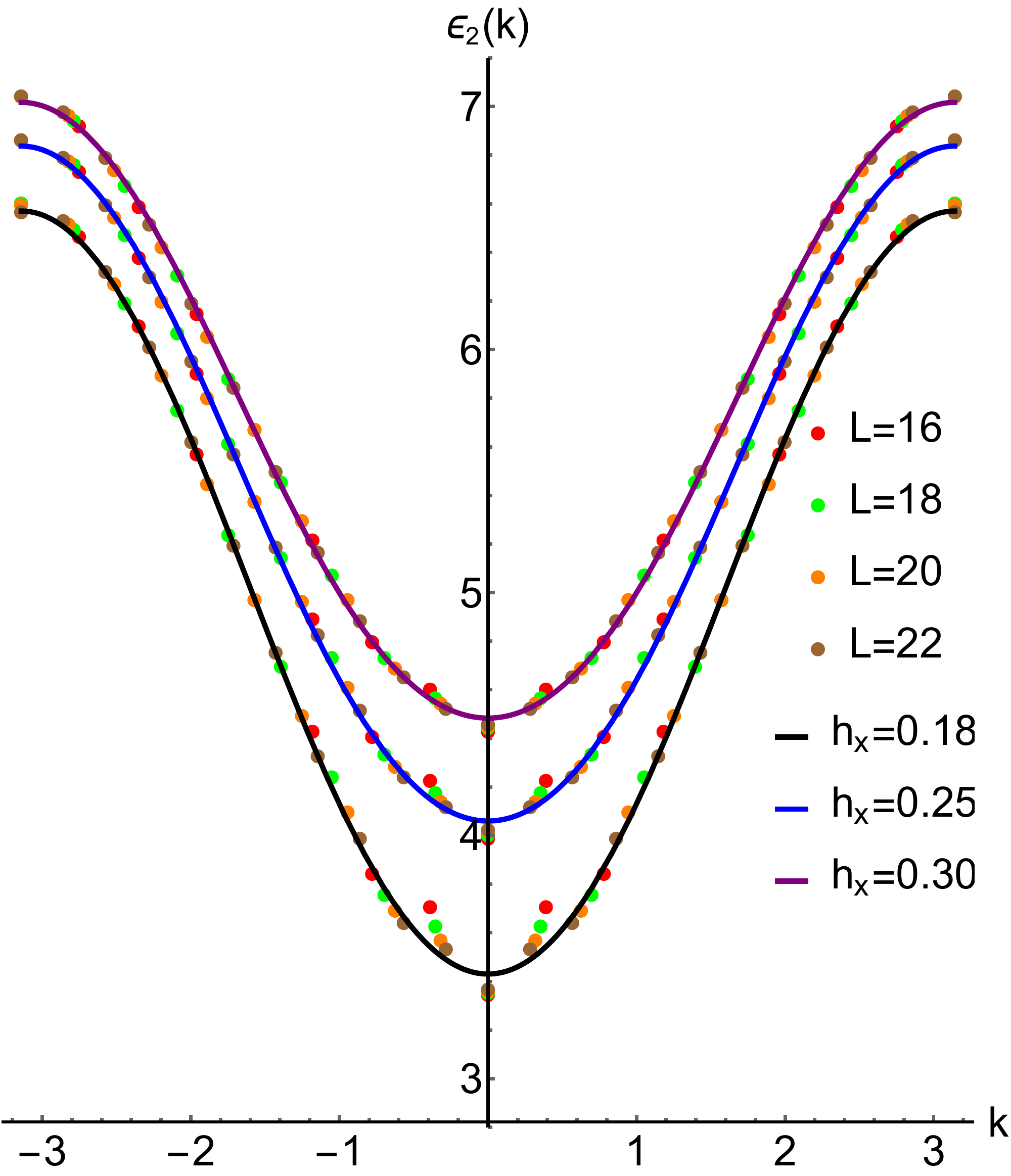} 
\par\end{centering}
\caption{\label{fig:The-second-quasi-particle} The second quasi-particle dispersion
relation for $h_{z}=1.5$ and $h_{x}=0.18$, $0.25$ and $0.30$.
The differently colored dots are energy levels computed for systems
sizes of $L=16$, $18$, $20$ and $22$ spins, illustrating that
finite size dependence is already negligible. The continuous lines
are fits of a function $\epsilon_{2}(k)=\sqrt{A+B\cos k+C\cos2k}$.}
\end{figure}

The numerically obtained dispersion relation $\epsilon_{2}(k)$ for
the second quasi-particle is shown in Fig. \ref{fig:The-second-quasi-particle}
for the case $h_{z}=1.5$ and $h_{x}>h_{x}^{(2)}$. Similarly to the
case of the fundamental excitation discussed in Subsection \ref{subsec:The-first-quasi-particle},
the quasi-particle gap (mass) increases with $h_{x}$, while the Lieb-Robinson
velocity 
\begin{equation}
v_{2\text{max}}=\max_{k}\frac{d\epsilon_{2}}{dk}
\end{equation}
decreases as shown in Table \ref{tab:Values-of-the-1}.
\begin{center}
\begin{table}
\begin{centering}
\begin{tabular}{|c|c|c|c|}
\hline 
$h_{x}$  & $0.18$  & $0.25$  & $0.30$\tabularnewline
\hline 
\hline 
$v_{2\text{max}}$  & $1.579$  & $1.413$  & $1.295$\tabularnewline
\hline 
\end{tabular}
\par\end{centering}
\caption{\label{tab:Values-of-the-1}Values of the Lieb-Robinson velocity determined
from the data shown in Fig. \ref{fig:The-second-quasi-particle}.}
\end{table}
\par\end{center}

It is also clear from Fig. \ref{fig:The-second-quasi-particle} that
the second quasi-particle mass depends more strongly on the chain
length $L$, especially when $h_{x}$ is closer to the threshold value
where the bound state appears. The reason is that the weaker the binding,
the larger is the spatial extension of the two-body wave-function,
therefore the more it is distorted in finite volume.

The numerical spectra of the spin chain show that the ratios $\Delta_{a1}$
are consistently higher than the continuum $E_{8}$ values and increase
with $h_{z}$. As a result, the third particle can be observed only
for the cases $h_{z}=1.25$ and $h_{z}=1.5$, where the critical values
can be obtained in a similar way as for the second particle, and turn
out to be $h_{x}^{(3)}\approx0.79$ and $h_{x}^{(3)}\approx1.82$.
In addition, the third particle is extremely loosely bound for all
values of $h_{x}$ where it exists, and the numerical data suggest
that it may eventually become unbound for much larger $h_{x}$ although
this is hard to nail down with sufficiently high precision due to
finite size effects. This explains why there is no signature of the
third particle in the entropy slope. Indeed, a quantum quench results
in a ``plasma'' of finite energy density, which destabilizes any
sufficiently loosely bound state by collisions with the particles
present. One still expects some weak resonance in the spectral density
of the two-particle continuum, though, and indeed hints of such a
resonance state can be seen in the power spectra discussed in Section
\ref{sec:Relation-to-the}.

\subsection{Post-quench quasi-particle density \label{subsec:Post-quench-quasi-particle-pictu}}

\begin{figure}
\subfloat[$h_{z}=1.25$]{\begin{centering}
\includegraphics[width=0.7\columnwidth]{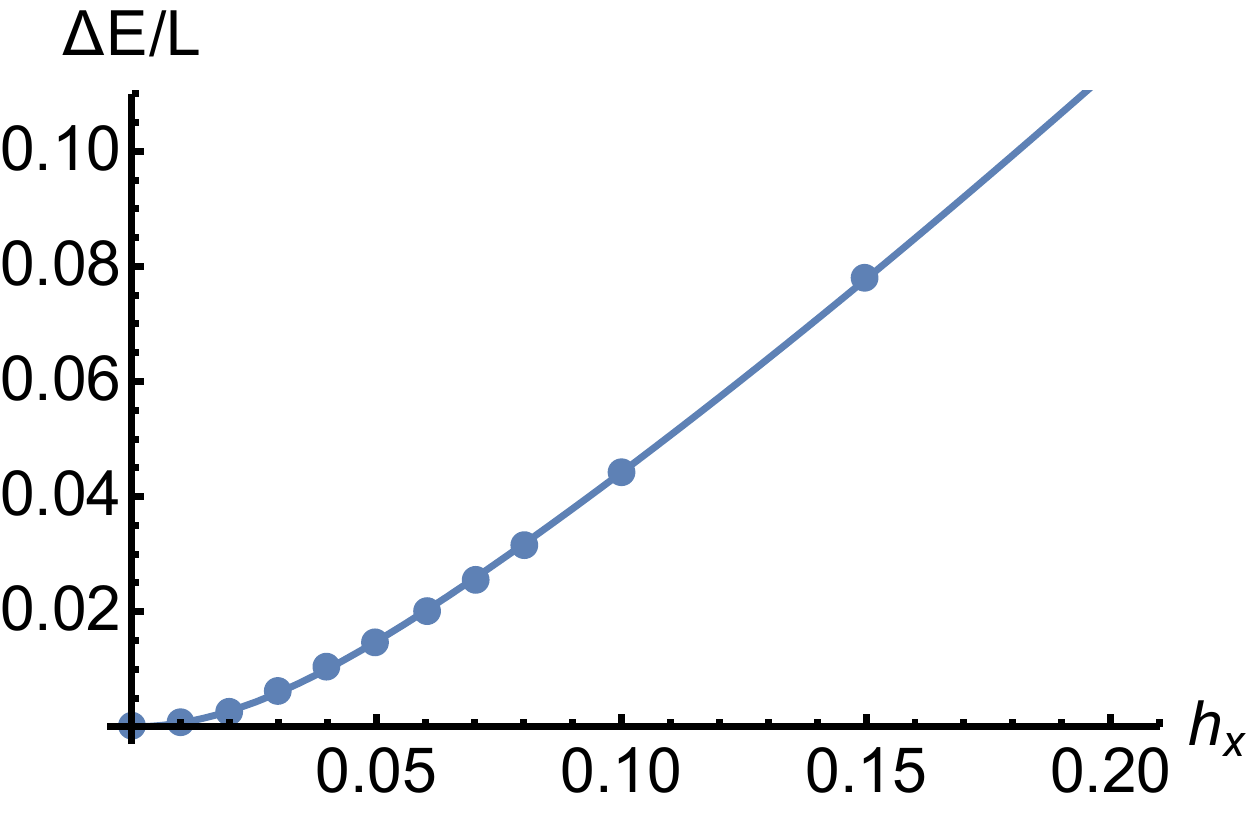} 
\par\end{centering}
}

\subfloat[$h_{z}=1.5$]{\begin{centering}
\includegraphics[width=0.7\columnwidth]{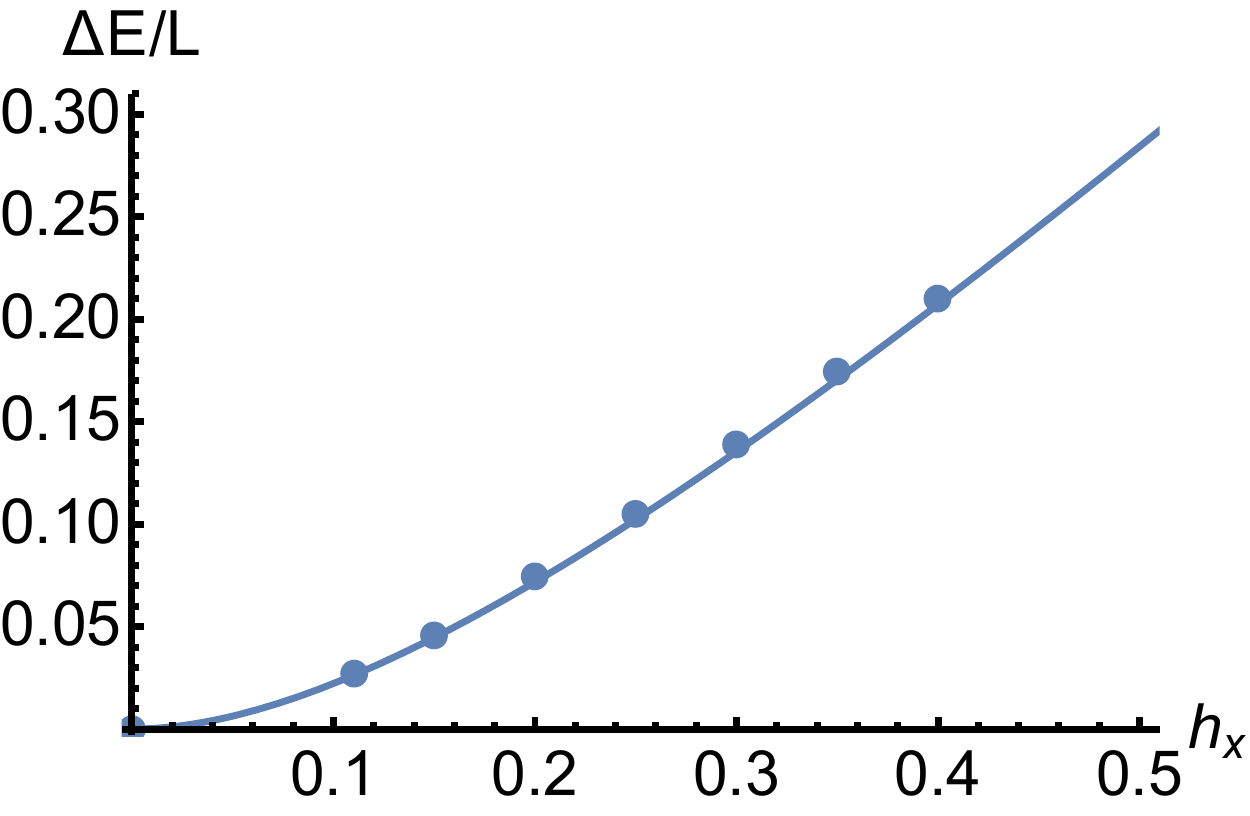} 
\par\end{centering}
}

\caption{\label{fig:The-energy-density} The energy density after a quantum
quench from $h_{x}=0$ to $h_{x}\protect\neq0$ for $h_{z}=1.25$
and $1.5$.}
\end{figure}

Finally, in Fig. \ref{fig:The-energy-density} we illustrate that
the quenches we consider have very low quasi-particle density. The
plots show the energy pumped into the quench, defined as the expectation
value of the post-quench Hamiltonian minus the post-quench ground
state eigenvalue, per lattice site (in units $J=1$). One can put
a simple upper bound on the particle density by dividing the energy
density with the value of the gap. For the critical value $h_{x}=0.04$
at $h_{z}=1.25$ the upper bound on the particle density is about
one particle per $70$ lattice sites, while for the critical value
$h_{x}=0.14$ at $h_{z}=1.5$ this results in a density of about one
particle per $35$ lattice sites. Even for the case $h_{x}=0.4$ at
$h_{z}=2$ the upper bound is one particle per $25$ lattice sites,
still a very low density compared to the correlation length $\xi$
which can be bounded from above by its value at $h_{x}=0$ \cite{pfeuty}
\begin{equation}
\xi=\frac{1}{\log|h_{z}|}=\begin{cases}
4.48 & h_{z}=1.25\\
2.47 & h_{z}=1.5\\
1.44 & h_{z}=2.0
\end{cases}
\end{equation}
given in number of lattice sites. This demonstrates that the post-quench
particle density is very small for the parameter range of interest.

\section{Relation to the Gibbs paradox\label{sec:Relation-to-the}}

In the following table the positions $h_{x}^{\text{min}}$ of the
minima of the mean entropy production rate $\overline{\partial_{t}S}$
(Table \ref{tab:Position-of-the}) are compared with the threshold
values $h_{x}^{(2)}$ where the second bound state appears (Table
\ref{tab:Critical-values-of})
\begin{center}
\begin{tabular}{|c|c|c|c|c|}
\hline 
$h_{z}$  & $1.25$  & $1.5$  & $1.75$  & $2$\tabularnewline
\hline 
\hline 
$h_{x}^{\text{min}}$  & $0.040$  & $0.140$  & $0.268$  & $0.412$\tabularnewline
\hline 
$h_{x}^{(2)}$ & $0.040$  & $0.146$  & $0.261$  & $0.400$\tabularnewline
\hline 
\end{tabular}
\par\end{center}

A crucial observation is that these values are very close: for smaller
values of $h_{z}$ they eventually coincide within numerical accuracy,
while for the higher values $h_{z}=1.75$ and $h_{z}=2$ the minimum
appears at a slightly larger $h_{x}$ than the bound state threshold.

For the interpretation of these results it is important to recall
first that at late times the asymptotic entanglement entropy of a
large subsystem can be interpreted as the thermodynamic entropy \cite{exp2,calcard_entropy,beugeling,deutsch,integrable_entanglement_growth}.
To understand the association between the bound states and the entropy
production rate, we turn to a quasi-particle description of entropy
production. The quasi-classical picture of quench dynamics \cite{calabrese-cardy}
describes the initial state as a source of entangled quasi-particle
pairs which propagate to different parts of the system, resulting
in the build-up of spatial correlations and entanglement growth. This
picture was explicitly demonstrated for integrable quenches in the
Ising spin chain \cite{calessfag} and also forms the basis of a semi-classical
approach for quantum quenches \cite{semicl}, which is expected to
be valid for sufficiently small post-quench density even in the non-integrable
case. It also successfully describes entropy production in integrable
systems \cite{integrable_entanglement_growth,multi-particle_case,entanglement_ghd}
and leads to the following formula for the late time growth of the
entanglement entropy of a subsystem of size $\ell$ \cite{calcard_entropy,integrable_entanglement_growth,multi-particle_case}:
\begin{equation}
S(t)\propto2t\sum_{n}\int_{2v_{n}t<\ell}dkv_{n}(k)f_{n}(k)+\ell\sum_{n}\int_{2v_{n}t>\ell}dkf_{n}(k)\;,\label{eq:semicl_entropy}
\end{equation}
where $n$ enumerates the different quasi-particle species, $k$ is
the momentum of the quasi-particles, $v_{n}(k)$ is their velocity
and $f_{n}(k)$ is a rate function describing the entropy produced
by quasi-particle pairs of species $n$ which depends on their production
rate. For the half-system ($\ell=\infty$) entanglement entropy the
second term describing saturation is absent, and the integral in the
first one has no restriction\footnote{The restriction in the integral leads to light-cone propagation as
a consequence of the Lieb–Robinson bound.} so it simplifies to 
\begin{equation}
S(t)\propto2t\sum_{n}\int dkv_{n}(k)f_{n}(k)\,.
\end{equation}

Eq. (\ref{eq:semicl_entropy}) suggests that the entanglement production
rate is a slowly varying function of the quench parameter $h_{x}$
and the data in Fig. \ref{fig:Top-diagrams:-time} show that this
is indeed true below the threshold $h_{x}^{(2)}$. Note that after
an initial rise, the contribution from the first species ($A_{1}$)
decreases which is explained below in terms of the quasi-particle
spectrum. If the effect of the new quasi-particle ($A_{2}$) simply
added the contribution of pairs $A_{2}A_{2}$, it should have the
same behavior as the contribution from pairs $A_{1}A_{1}$, except
being smaller due to the even larger gap and smaller quasi-particle
velocity.

However, as demonstrated in Fig. \ref{fig:Top-diagrams:-time}, the
entanglement production rate increases by an order of magnitude after
passing the threshold, an effect which is really pronounced closer
to the critical point $h_{z}=1$. The flaw in the naive argument is
that it neglects the contribution of species mixing, which is the
cornerstone of the classical Gibbs paradox. In the usual setting of
the paradox one takes a box divided by a wall into two equal halves,
with $N$ particles in each. Even though removing the wall is reversible
by reinserting it, a simple computation using ideal gas laws shows
that it increases the thermodynamic entropy by an amount $\Delta S=2k_{B}N\ln2$.
The key to resolving the paradox is to specify the relation between
particles in the two halves: for indistinguishable particles, this
term is not present, while if the particles are distinguishable, it
corresponds to their mixing entropy and removing the wall is indeed
an irreversible process.

Similarly, the appearance of the second quasi-particle increases the
thermodynamic entropy produced in the quench by the species information.
This is supported by the finding that in the continuum limit of the
Ising spin chain, quenching in $h_{x}$ results in the creation of
mixed pairs $A_{1}A_{2}$ \cite{e8_quenches}. The presence of mixed
pairs means that the entropy carried by the quasi-particles contains
species information, i.e. the Gibbs mixing entropy. In Appendix \ref{sec:A-semiclassical-calculation}
we demonstrate via a semiclassical estimate using features of the
pair amplitudes from the field theory and a construction recently
developed in \cite{entanglement_ghd}, that the mixed pairs indeed
lead to an increase of roughly the observed magnitude in the entropy
production rate. 

\begin{figure*}
\begin{centering}
\includegraphics[height=10cm]{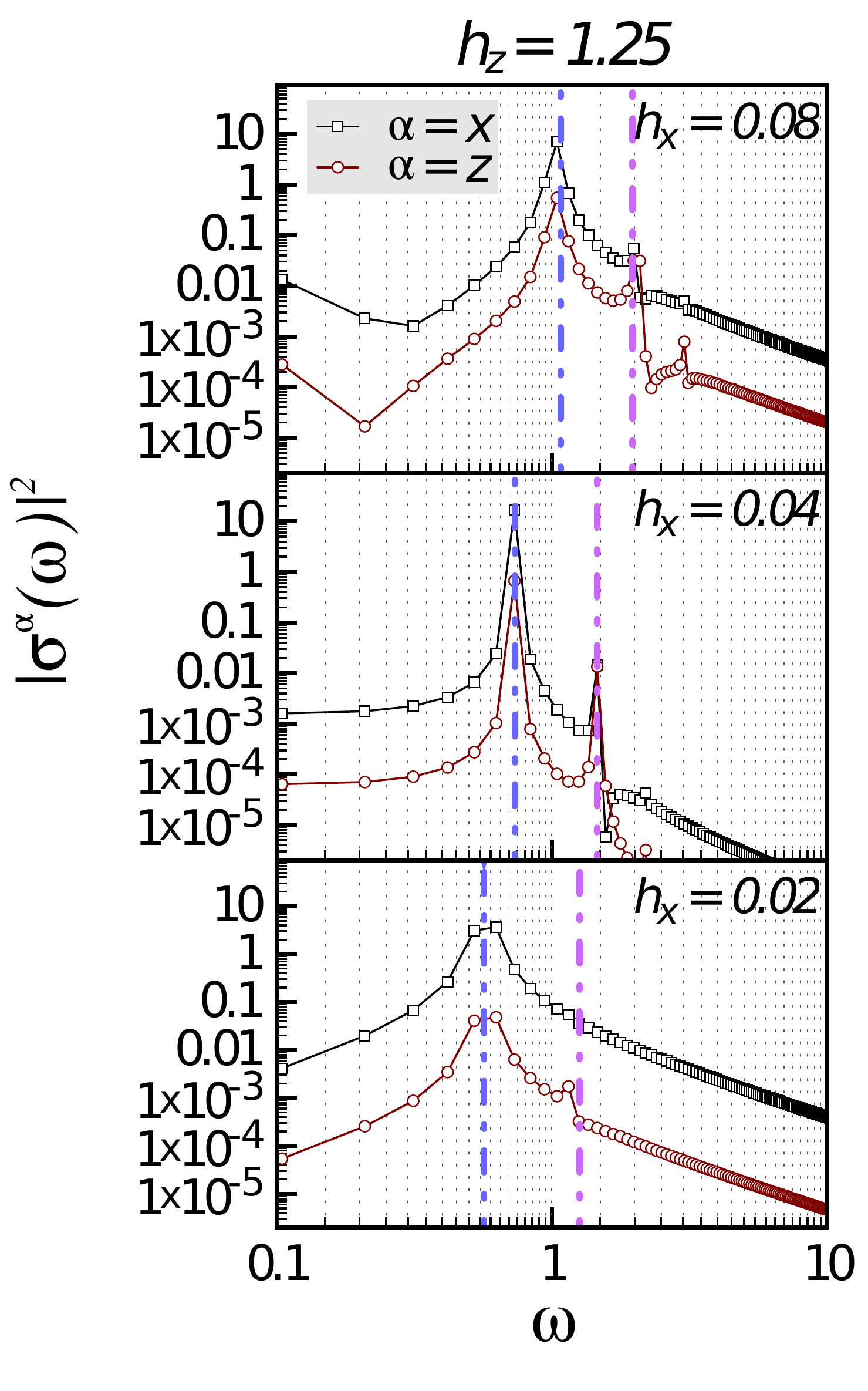}~~~~~~~\includegraphics[height=10cm]{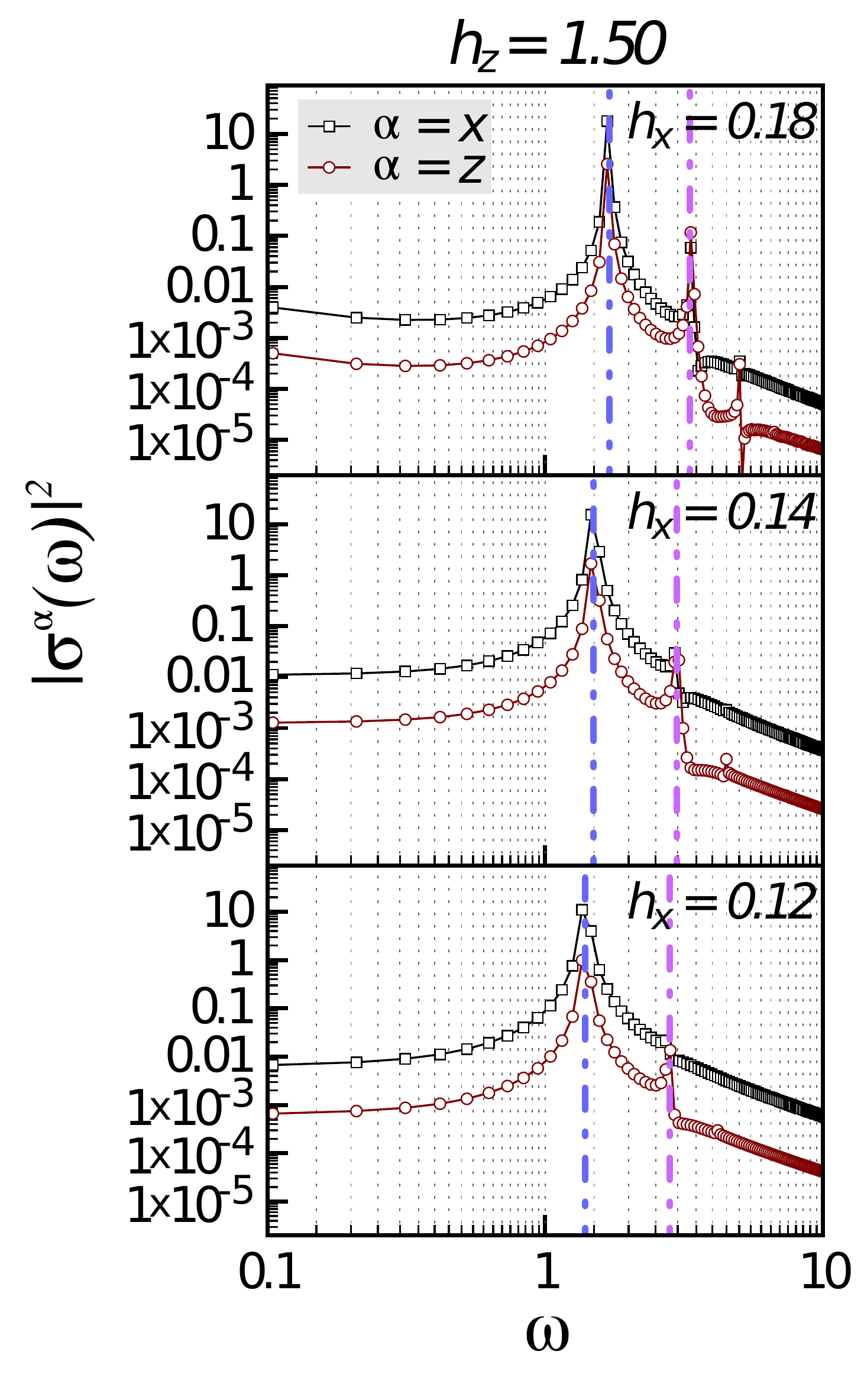} 
\par\end{centering}
\caption{\label{fig:Fourier-transforms-of} Fourier transforms of the time
dependence of longitudinal ($\alpha=x$) and transverse ($\alpha=z$)
magnetizations. For each $h_{z}$ the three plots shown are before/around/beyond
the critical value $h_{x}^{\text{crit}}$. The (blue/purple) vertical
dash-dotted lines are the expected positions of the peaks corresponding
to the first and second quasi-particles $A_{1}$ and $A_{2}$. For
the bottom plots which are below threshold, the second vertical line
corresponds to the energy of the lowest lying two-particle state,
which turns into a zero-momentum $A_{2}$ state for $h>h_{x}^{(2)}$.
Note that for large enough $h_{x}$ a third peak emerges in the spectrum,
which is the precursor of the third quasi-particle $A_{3}$ discussed
in Subsection \ref{subsec:The-bound-state}. }
\end{figure*}

It is important to realize that in spite of the non-integrability
of the system, the quasi-particle picture is still expected to be
a good approximation. The reason is that turning on a longitudinal
field in the paramagnetic regime does not lead to a drastic change
in the physical behavior contrary to the ferromagnetic case \cite{Ising_confinement},
where it triggers confinement \cite{mccoy_wu}. Fig. \ref{fig:Fourier-transforms-of}
presents power spectra obtained from 
\begin{equation}
\sigma^{\alpha}(\omega)=\int_{0}^{\infty}dte^{i\omega t}\left\langle \sigma^{\alpha}(t)\right\rangle \;,
\end{equation}
where $\left\langle \sigma^{\alpha}(t)\right\rangle $ ($\alpha=x,z$)
are the longitudinal and transverse magnetizations. These show clear
quasi-particle peaks at the frequencies predicted by the exact diagonalization
results in Section  \ref{sec:Quasi-particle-spectrum-of}. In addition,
the self-consistency of the quasi-particle description is also demonstrated
by the small values of the upper limits of the post-quench particle
density obtained in Subsection \ref{subsec:Post-quench-quasi-particle-pictu}.

To explain the decrease of the entropy production rate below $h_{x}^{(2)}$
seen in Fig. \ref{fig:Top-diagrams:-time}, note that both the exact
diagonalization results (Fig. \ref{fig:The-gap-and}) and the power
spectra (Fig. \ref{fig:Fourier-transforms-of}) show that the particle
masses (excitation gaps) increase with $h_{x}$. Even though the post-quench
energy density $(\langle\Psi_{0}|H|\Psi_{0}\rangle-E_{0})/L$ increases
with $h_{x}$, its ratio with the energy gap saturates, which is also
consistent with the stagnation of the size of the quasi-particle peaks
in Fig. \ref{fig:Fourier-transforms-of}. This gives a (very rough)
upper bound on the particle density in the initial state, and so the
rate functions $f_{n}(k)$ (while not directly accessible) are also
expected to stop growing with $h_{x}$. Moreover, only a small fraction
of the quasi-particle excitations propagates at the maximum velocity;
this fact, joined with the global decrease (for all momenta) of the
quasi-particle velocities $v_{n}(k)$ with $h_{x}$ explains why the
late time mean entropy production rate $\overline{\partial_{t}S}=2{\displaystyle \sum_{n}}\int dkv_{n}(k)f_{n}(k)$
decreases for $h_{x}<h_{x}^{(2)}$. 

Albeit the trend change in $\overline{\partial_{t}S}$ as a function
of $h_{x}$ is rapid, it is not a discontinuous jump due to several
reasons. First, the heavier second excitation is produced with a density
that smoothly depends on the quench parameter $h_{x}$ and increases
only gradually. Second, the distinguishability of the second quasi-particle
peak also increases gradually with $h_{x}$. As shown by the power
spectra in Fig. \ref{fig:Fourier-transforms-of}, at first the second
quasi-particle peak is not prominent and is barely distinguishable
from the continuum background. As known in the case of the equilibrium
Gibbs paradox \cite{gibbs1,gibbs2}, distinguishability is a key feature
governing the effective number of species contributing to thermodynamic
quantities such as free energy and entropy. Third, the post-quench
system is filled with a finite density ``plasma'' of excitations
which leads to a finite life-time of the quasi-particle excitations,
and is also known to lead a shift in the effective quasi-particle
masses \cite{cubero_schuricht}. In case of very weakly bound quasi-particles
(such as the third quasi-particle which does exist at zero temperature/density
for suitably large $h_{x}$), the plasma effect can even suppress
the signal completely by destabilizing the excitations. This effect
is completely consistent with, and indeed explains, the observation
that the difference between $h_{x}^{(2)}$ and $h_{x}^{\mathrm{min}}$
grows with increasing $h_{z}$.

As a consequence of the gradual change of the effective number of
quasi-particle species characterizing the post-quench state, the simple
summation over quasi-particle species appearing in Eq. (\ref{eq:semicl_entropy})
does not eventually apply in the region around the threshold. Therefore
a quantitative explanation of $\overline{\partial_{t}S}$ as a function
of $h_{x}$ requires a more complete theory of entropy production
with multiple quasi-particle species after a non-integrable quench,
which at this point is left open for the future. While this affects
the exact definition of the rate functions $f_{n}(k)$, it is not
expected to alter the relation between the asymptotic entropy density
and entanglement production rate (the two terms of Eq. (\ref{eq:semicl_entropy}))
which is a general consequence of the quasi-particle picture alone.

\section{Discussion}

In this paper we found an anomalous increase of the entropy production
rate due to the appearance of bound states in the quantum Ising spin
chain quenched by switching on a longitudinal magnetic field at a
fixed value of the transverse field in the paramagnetic phase. The
anomaly is clearly related to the appearance of a new quasi-particle
state in the spectrum, and its details confirm that the effect is
a dynamical manifestation of the Gibbs paradox well-known from equilibrium
statistical mechanics. We remark that after the completion of this
work, new results obtained for the $3$-state Potts spin chain show
exactly the same behavior as reported here for the Ising chain; details
will be published elsewhere \cite{potts}.

We emphasize that there is a crucial difference between integrable
systems, where the effect of multiple species on entropy production
is simply described by the summation in (\ref{eq:semicl_entropy})
(cf. Ref. \cite{multi-particle_case}), and the non-integrable case
considered here. In integrable systems there exist infinitely many
conserved charges which distinguish the quasi-particle excitations,
forcing them to be absolutely stable and their scattering to be completely
elastic. However, in our case integrability is broken by the longitudinal
field, and no charges differentiate between the quasi-particles; their
separate identity depends on distinguishability of the corresponding
spectral peaks \cite{gibbs2}, which any full theory of entanglement
production in non-integrable systems must inevitably take into account.
However, even without such a detailed formalism it is absolutely clear
that mixing entropy has a large effect on the entropy density of the
steady state according to our understanding of the equilibrium Gibbs
paradox.
\begin{acknowledgments}
The authors are grateful to P. Calabrese for invaluable comments and
suggestions on the draft. M.K. and G.T. also thank R. Moessner and
B. Dóra for useful discussions and comments. This research was supported
by the National Research Development and Innovation Office (NKFIH)
under a K-2016 grant no. 119204, and also by the BME-Nanotechnology
FIKP grant of EMMI (BME FIKP-NAT). M.C. acknowledges support by the
Marie Sklodowska-Curie Grant No. 701221 NET4IQ, M.K. by a ``Prémium”
postdoctoral grant of the Hungarian Academy of Sciences, while G.T.
was also supported by the Quantum Technology National Excellence Program
(Project No. 2017-1.2.1-NKP-2017- 00001). The authors also acknowledge
the hospitality of the Erwin Schrödinger Institute (Vienna) while
working on the manuscript.
\end{acknowledgments}

\appendix

\section{A semiclassical calculation\label{sec:A-semiclassical-calculation}}

To illustrate the mechanism behind the increase of entanglement growth,
in this appendix we present a semiclassical calculation in systems
with two distinguishable species of free fermions. We follow and extend
the method introduced in the recent work \cite{BFPC}. On the one
hand, the fermionic algebra considerably simplifies the calculation,
on the other hand it serves as a first approximation to the system
studied in this paper insofar as the lightest particle is a genuine
free fermion for $h_{x}=0.$

\begin{figure*}[t!]
\includegraphics[width=0.45\textwidth]{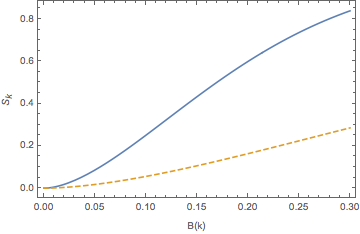} \hfill{}\includegraphics[width=0.45\textwidth]{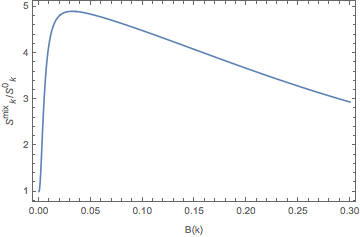}
\caption{\label{fig:Sk} Entanglement contribution of the $\{k,-k\}$ sector
as a function of the amplitude $B(k)$ with $A(k)=0.005$ held fixed.
\emph{Left panel:} The lower dashed curve represents the no-mixing
case $S_{k}^{0}$ where $C(K)=D(k)=0$, while the upper solid curve
shows the result $S_{k}^{\text{mix}}$ in the presence of mixed pairs,
$C(k)=D(k)=1.6B(k).$ \emph{Right panel:} Ratio $S_{k}^{\text{mix}}/S_{k}^{0}$
of the entropies in the mixing and non-mixing case.}
\end{figure*}
We assume that during the quench entangled pairs of quasi-particles
are created with opposite momenta. The initial state (in a finite
volume) thus can be written as 
\begin{multline}
|\Psi_{0}\rangle=\mathcal{N}\prod_{k>0}\left[1+A(k)a_{k}^{\dagger}a_{-k}^{\dagger}+B(k)b_{k}^{\dagger}b_{-k}^{\dagger}\right.\\
\left.+C(k)a_{k}^{\dagger}b_{-k}^{\dagger}+D(k)b_{k}^{\dagger}a_{-k}^{\dagger}\right]|0\rangle\,,\label{psi0}
\end{multline}
where $|0\rangle$ is the post-quench ground state, $\mathcal{N}$
is a normalization factor, and $a_{k}^{(\dagger)}$ and $b_{k}^{(\dagger)}$
are the annihilation (creation) operators of the first and the second
particle, respectively, obeying anticommutation relations 
\begin{equation}
\{a_{p},a_{q}^{^{\dagger}}\}=\{b_{p},b_{q}^{\dagger}\}=\delta_{p,q}\,,\quad\{a_{p},b_{q}\}=\{a_{p},b_{q}^{\dagger}\}=0\,.
\end{equation}
 The product in Eq. \eqref{psi0} runs over positive momenta quantized
in a finite volume $L.$ Note that we allow for the creation of mixed
pairs consisting of an $a$-type and a $b$-type particle. Using the
fermionic algebra one can compute the normalization factor with the
result 
\begin{multline}
\mathcal{N}=\prod_{k>0}\mathcal{N}_{k}\\
=\prod_{k>0}\left(1+|A(k)|^{2}+|B(k)|^{2}+|C(k)|^{2}+|D(k)|^{2}\right)^{-1/2}\,.
\end{multline}
The density matrix also factorizes into momentum sectors,
\begin{equation}
\hat{\rho}_{0}=|\Psi_{0}\rangle\langle\Psi_{0}|=\prod_{k>0}\hat{\rho}{}_{k,-k}\,.
\end{equation}

The idea behind the semiclassical picture for entanglement generation
is that a spatial subsystem becomes entangled with the rest of the
system via the entanglement of particle pairs for which one member
of the pair is inside the subsystem while the other member is outside
of it \cite{CC}. Each momentum sector thus contributes by the entanglement
entropy between the two modes of momentum $k$ and $-k$, so we need
to compute the reduced density matrices 
\begin{equation}
\hat{\rho}_{k}=\mathrm{Tr}_{-k}\,\hat{\rho}_{k,-k}\,.
\end{equation}
Due to the fermionic nature of particles, the mode $-k$ corresponds
to a 4-dimensional space with the basis 
\begin{equation}
|0\rangle_{-k}\,,\quad a_{-k}^{\dagger}|0\rangle_{-k}\,,\quad b_{-k}^{\dagger}|0\rangle_{-k}\,,\quad a_{-k}^{\dagger}b_{-k}^{\dagger}|0\rangle_{-k}\,.
\end{equation}
Taking the trace of $\hat{\rho}_{k,-k}$ in this basis we arrive at 
\begin{widetext}
\begin{multline}
\hat{\rho}_{k}=\mathcal{N}_{k}^{2}\left[|0\rangle\langle0|+\left(|A(k)|^{2}+|C(k)|^{2}\right)a_{k}^{\dagger}|0\rangle\langle0|a_{k}+\left(|B(k)|^{2}+|D(k)|^{2}\right)b_{k}^{\dagger}|0\rangle\langle0|b_{k}\right.\\
+\left.\left[A(k)D(k)^{*}+B(k)^{*}C(k)\right]a_{k}^{\dagger}|0\rangle\langle0|b_{k}+\left[A(k)*D(k)^{*}+B(k)C(k)^{*}\right]b_{k}^{\dagger}|0\rangle\langle0|a_{k}\right]\,,
\end{multline}
\end{widetext}

\noindent where we dropped the $-k$ subscript from the Fock vacuum
state $|0\rangle.$ The corresponding entanglement entropy is then
\begin{equation}
S_{k}=-\mathrm{Tr}_{k}\,\hat{\rho}_{k}\log\hat{\rho}_{k}\,.
\end{equation}
At time $t$ only those pairs contribute to the half space entanglement
entropy that come from the $[-v_{k}t,v_{k}t]$ interval, which in
the infinite volume limit leads to 
\begin{equation}
S(t)=-\int\frac{dk}{2\pi}2v_{k}t\,\mathrm{Tr}_{k}\,\hat{\rho}_{k}\log\hat{\rho}_{k}\,,
\end{equation}
 an entanglement entropy growing linearly in time.

Let us analyze how the contribution $S_{k}$ of the $\{k,-k\}$ sector
is affected by the presence of mixed pairs. $S_{k}$ depends on the
four amplitudes which we fix using the numerical values that were
measured in Ref. \cite{HKT} for a similar quench in the continuum
Ising field theory (see Fig. 5.4 there). In particular, we set $A(k)=0.005,$
$C(k)=D(k)$ due to parity symmetry, and a relation between $B(k)$
and the mixing amplitudes: $C(k)=D(k)=1.6B(k).$ Keeping $A(k)$ fixed
is a meaningful choice because we are interested in the change of
the entanglement production rate around the threshold for the second
particle, where $B(k)$ starts to grow from zero but $A(k)$ is approximately
constant.

In the left panel of Fig. \ref{fig:Sk} we plot $S_{k}$ both in the
presence (solid curve) and in the absence ($C(k)=D(k)=0$) of mixed
pairs (dashed curve) in the initial state as a function of the creation
amplitude $B(k)$ of the second particle. It is clear that in accordance
with the Gibbs mixing entropy, the presence of mixed pairs leads to
an enhancement of the entanglement entropy and of the entanglement
generation rate. In the right panel the ratio of the two curves are
plotted demonstrating that passing the threshold there is a sudden
and significant increase in the entanglement entropy as a result of
the mixed pairs.

\section{\label{sec:Numerical-simulation-of} Numerical simulation of time
evolution}

Numerical simulations of the quench dynamics in the non-integrable
Ising chain was performed using the infinite volume Time-Evolving
Block-Decimation (iTEBD) algorithm \cite{vidal1}. The algorithm exploits
the translational invariance of the system by representing a generic
many-body state on a one-dimensional lattice as 
\begin{equation}
|\Psi\rangle=\sum_{\ldots,s_{j},s_{j+1},\ldots}\cdots\Lambda_{o}\Gamma_{o}^{s_{j}}\Lambda_{e}\Gamma_{e}^{s_{j+1}}\cdots|\ldots,s_{j},s_{j+1},\ldots\rangle\;,
\end{equation}
where $s_{j}$ spans the local spin-$1/2$ Hilbert space, $\Gamma_{o/e}^{s}$
are $\chi\times\chi$ matrices associated with the odd/even lattice
site; $\Lambda_{o/e}$ are diagonal $\chi\times\chi$ matrices with
the singular values corresponding to the bipartition of the system
at the odd/even bond as their entries. The many-body state is initialized
to the product state $|\Psi_{0}\rangle=\bigotimes(|\!\!\uparrow\rangle+|\!\!\downarrow\rangle)/\sqrt{2}$.

The Matrix Product State (MPS) representation of the ground state
$|\Psi_{GS}\rangle$ is obtained by time-evolving the initial state
$|\Psi_{0}\rangle$ in imaginary time. We used a second-order Suzuki-Trotter
decomposition of the evolution operator with imaginary time Trotter
step $\tau=10^{-4}$. The Hamiltonian was been tuned to the paramagnetic
phase of the model, namely $h_{x}=0$ and $h_{z}\in\{1.25,1.5,1.75,2\}$.
Due to the presence of an energy gap separating the ground state from
the rest of the spectrum, an auxiliary dimension $\chi_{0}=32$ was
sufficient to have a very accurate MPS description of the ground state.

Similarly, the post-quench time evolution was obtained by evolving
the corresponding ground state with a new Hamiltonian with $h_{x}\neq0$
in real time. For this purpose again a second-order Suzuki-Trotter
decomposition of the evolution operator was used, with real time Trotter
step $dt=10^{-3}$. In order to keep the truncation error as small
as possible, the auxiliary dimension was allowed to grow up to $\chi_{MAX}=512$
which was sufficient to reach a maximum time $T=60$. The ability
to reach relatively large times is related to the dynamical properties
of the system under investigation. As explained in the main text,
for such class of quenches, the bipartite entanglement entropy does
not growth significantly as long as $h_{x}$ is ``sufficiently''
small. For $h_{x}$ larger than the critical threshold, the bipartite
entanglement entropy starts growing faster, nonetheless always remaining
smaller than $\simeq3$. After a relatively short transient, the numerical
data for the entanglement entropy showed a linear increase (apart
from oscillations) whose slope depends on the particular value of
the longitudinal field exactly as expected after a global quantum
quench. In particular, a numerical estimation of the entanglement
entropy slope $\overline{\partial_{t}S}$ has been obtained by performing
a linear fit of the iTEBD data in the time-window $30\leq t\leq60$
(cf. Fig. \ref{fig:Top-diagrams:-time}).

Similarly, the iTEBD simulation allows us to trace the expectation
value of local observables easily. In particular, we analyzed the
longitudinal $\langle\sigma^{x}(t)\rangle$ and transverse $\langle\sigma^{z}(t)\rangle$
magnetizations. From the corresponding time series, the power spectra
$\sigma^{x/z}(\omega)$ were obtained using FFT (see Fig. \ref{fig:Fourier-transforms-of}),
with an angular frequency resolution $d\omega=2\pi/T\simeq0.10472$.
The second peak in the power spectrum which appears above the critical
value of $h_{x}$ is the signature of a new bound state, in agreement
with the predicted spectrum from exact diagonalization.
\end{document}